\documentstyle[11pt,newpasp,twoside,epsf]{article}
\def\Msun{$M_{\sun}$}

\def\yr {yr$^{-1}$}

\def\kms{km s$^{-1}$}

\def\edcomment#1{\iffalse\marginpar{\raggedright\sl#1\/}\else\relax\fi}
\reversemarginpar

\begin{document}
\title{Molecular Gas Properties in the Central Kiloparsec of Barred and Unbarred Spirals}
 \author{Kartik Sheth}
\affil{California Institute of Technology, MS 105-24, Pasadena, CA 91125 \\ \& University of Maryland, College Park, MD 20742-2421}

\begin{abstract}
We study the molecular gas properties in the central kiloparsec of 29
barred and 15 unbarred spirals from the
BIMA\footnote{Berkeley-Illinois-Maryland Association} Survey of Nearby
Galaxies (SONG).  We find that the mean nuclear molecular gas surface
density of barred spirals ($\langle$$\Sigma_{nuc}$$\rangle$=309$\pm$71
\Msun\ pc$^{-2}$) is three times higher than that of unbarred spirals
($\langle$$\Sigma_{nuc}$$\rangle$=107$\pm$29 \Msun\ pc$^{-2}$).  Nine
out of the eleven bars with $\Sigma_{nuc} >$ 300 \Msun\ pc$^{-2}$ are
early types. Comparison with estimates of the star formation threshold
density ($\Sigma_{crit}$) indicates that enhanced star formation in
bars may be due to a larger fraction having $\Sigma_{nuc} > typical\
\Sigma_{crit}$. We also find that barred spirals are more centrally
concentrated than unbarred spirals.  The median value of the
concentration parameter for barred spirals
(f$_{con}$=$\Sigma_{nuc}$/$\Sigma_{disk}$=27.7) is a factor of four
higher than that for the unbarred spirals (f$_{con}$=6.2).  Finally we
investigate the dependence of the central concentration on bar
properties and find it may be weakly correlated with the bulge size,
but is not correlated with the bar length.  This suggests that inner
Lindblad resonances, bar ellipticity and circumnuclear star formation
play important roles in determining the gas accretion in the central
kiloparsec. \\

\noindent See http://www.astro.caltech.edu/$\sim$kartik/thesis.html
for more detailed description of these thesis results.  This research
was made possible with funding from NSF grants AST-9981308,
AST-9981289 and AST-9981546.
\end{abstract}

\section{Introduction}
Bars play a key role in galaxy evolution by inducing dramatic changes
 on relatively short time scales ($\sim$10$^8$ years).  Bar-induced
 gas mixing is known to reduce the overall metallicity gradient (e.g.,
 Martin \& Roy 1994).  Three decades of observational studies show
 that circumnuclear starburst activity is associated with bars (e.g.,
 S\'ersic \& Pastoriza 1965; Hawarden et al. 1986; Ho, Fillipenko \&
 Sargent 1997).  Models suggest that bars may be effective mechanisms
 for feeding active galactic nuclei (Shlosman, Frank \& Begelman
 1988), forming new bulges, and for evolving late Hubble type galaxies
 into earlier types (e.g., Friedli \& Benz 1993; Norman, Sellwood \&
 Hasan 1996).  The most dramatic prediction of models is that a bar
 can destroy itself if sufficient mass accretes ($\sim$
 1-2\%M$_{disk}$) in the center.  Since these evolutionary changes
 depend not only on the gas inflow but on the gas {\em accretion}, we
 compare the gas properties of barred and unbarred spirals in the
 central kiloparsec (kpc); we focus on the molecular gas because it is
 the dominant component of the ISM in inner regions of spirals
 (Scoville 1990).

Though good evidence for inflowing gas has been presented (e.g.,
Regan, Sheth \& Vogel 1999), only one study has examined whether the
inflowing gas accumulates in the center (OVRO-NRO study, Sakamoto et
al. 1999).  This twenty galaxy sample, however, pre-selected centrally
concentrated galaxies because of a CO-brightness selection criterion.
Moreover, it contained only 4 late type spirals preventing comparative
studies across Hubble type.  Hence a study with a larger and more
uniform sample is warranted.  The BIMA SONG provides such a database
(Regan et al. 2001).  SONG consists of 29 barred spirals and 15
unbarred spirals selected as follows: $\delta >$ --20$^o$, $i <$
70$^o$, V$_{HEL} <$ 2000 \kms, and perhaps most importantly, m$_B <$
11.

\begin{figure}
\vspace{-0.2in}
\plottwo{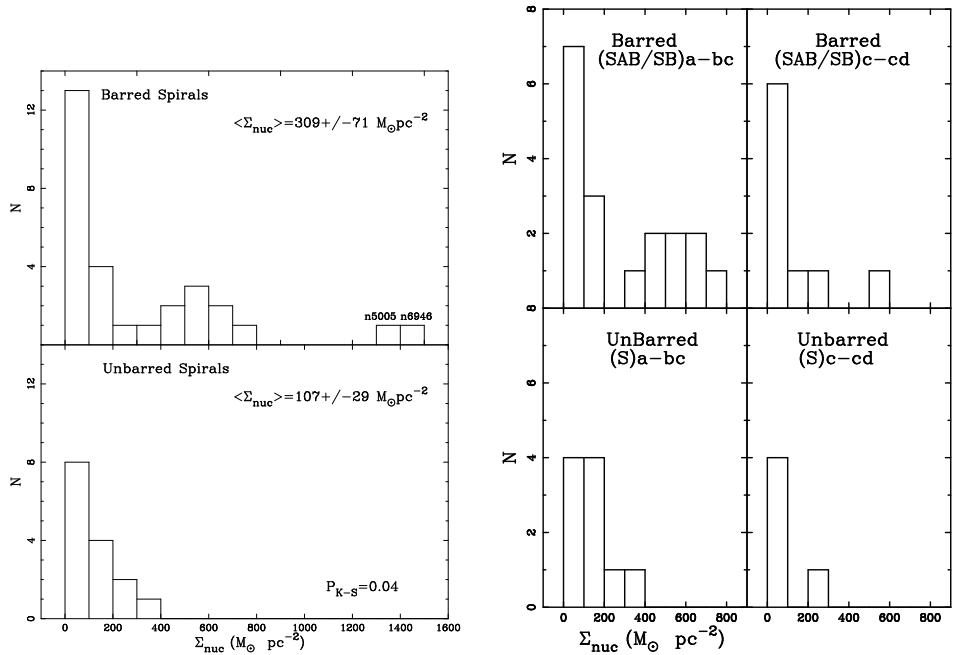}{signucbytype.talk.ps}
\caption{{\bf Left:} Histogram of $\Sigma_{nuc}$ in barred and
unbarred spirals from SONG.  Note the tail of high gas surface
densities in the barred spirals.  {\bf Right:} Same data now split
between early (Sa-Sbc) and late type (Sc-Sd) spirals.  Note that the 
highest gas surface densities are typically found in early type bars.}
\vspace{-0.2in}
\end{figure}

\section{Differences in Molecular Gas Surface Density}
In the left panel of Figure 1 we show the distribution of the nuclear
gas surface densities ($\Sigma_{nuc}$) in the central kpc diameter
region for all the SONG galaxies.  The mean $\Sigma_{nuc}$ for the
barred spirals ($\Sigma_{nuc}$=309.1$\pm$71.3 \Msun\ pc$^{-2}$) is
three times higher than that for the unbarred spirals
($\Sigma_{nuc}$=106.5$\pm$29.3 \Msun\ pc$^{-2}$).  A Kolmogrov-Smirnov
test shows that the distributions differ at a 96\% confidence level.
In the right panel, we show the same data split into early and late
type spirals. We find that the majority of the highest $\Sigma_{nuc}$
occur in early type galaxies. 9/11 bars with $\Sigma_{nuc} > $300 are
early types.

These differences in $\Sigma_{nuc}$ are important for understanding
circumnuclear star formation activity.  Using a simple gravitational
instability model (e.g., Kennicutt 1989), or a magneto-Jean's
instability model (e.g., Kim \& Ostriker 2001), we estimate that the
lower limit to the typical threshold or critical gas surface density
($\Sigma_{crit}$) is $\sim$100 \Msun\ pc$^{-2}$, assuming the
shallowest rotation curve from Sofue et al. (1999) and lowest velocity
dispersion from Galactic center measurements (Bally et al. 1988)
Though $\Sigma_{crit}$ undoubtedly varies from galaxy to galaxy
depending on the steepness of the rotation curve and the velocity
dispersion, 100 \Msun\ pc$^{-2}$ is a reasonable lower limit.

\begin{figure}
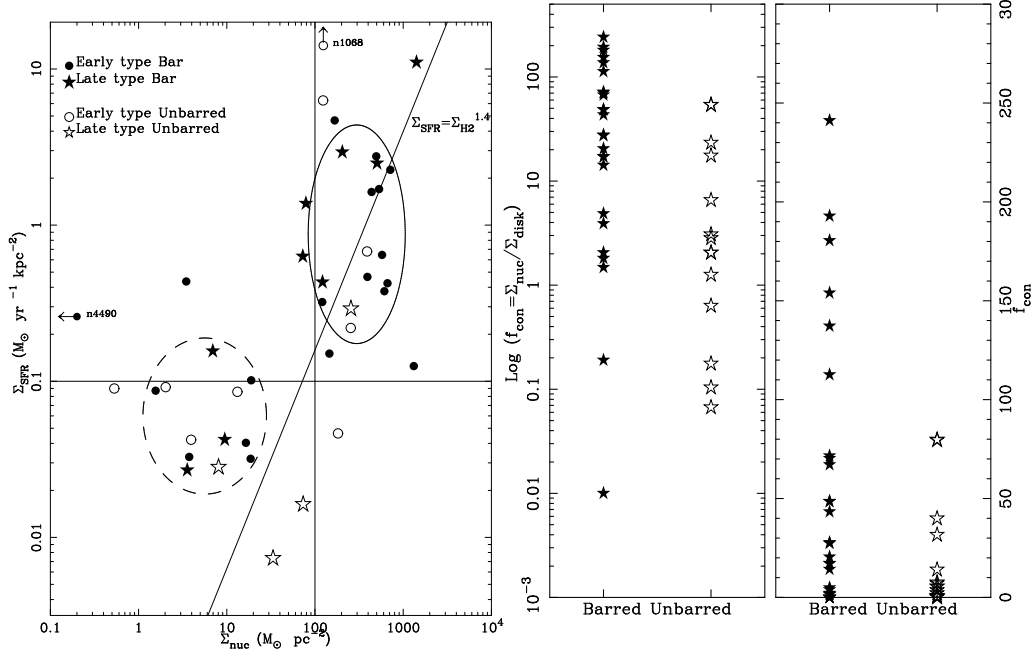

\vspace{0.9in}
\plotfiddle{sfschmidt.ps}{2.5in}{0}{37}{37}{-200}{0}
\plotfiddle{cenconc2.ps}{2.5in}{0}{37}{37}{-10}{200}
\vspace{-3.1in}
\caption{{\bf Left:} $\Sigma_{SFR}$ is plotted against $\Sigma_{nuc}$
for the SONG sample. The solid oval shows the high star formation
rates in galaxies above the threshold density (solid vertical line)
and the dashed oval shows the galaxies below the threshold.  The
diagonal line is the Schmidt law from Kennicutt (1998). {\bf Right:}
f$_{con}$ in barred and unbarred spirals is shown on a logarithmic
(left hand box) and a linear scale (right hand box); the two
distributions are distinctly different (P$_{K-S}$=0.07).  This is
strong statistical evidence for bar-induced gas inflow and accretion.}
\end{figure}

From Figure 1, we see that about half of the barred (16/29 or 55\%)
and unbarred (7/15 or 45\%) spirals have $\Sigma_{nuc} >$ 100 \Msun\
pc$^{-2}$.  But the highest gas surface densities in the unbarred
spirals are mostly between 100--200 \Msun\ pc$^{-2}$.  In fact, for
$\Sigma_{nuc} >$ 300 \Msun\ pc$^{-2}$, there is only 1/15 unbarred
spiral, compared to 11/29 barred spirals.  Considering that the
threshold density is likely to be higher than 100 \Msun pc${-2}$, and
in light of over three decades of studies indicating that bars, and
particularly {\em early type} bars, host circumnuclear starbursts, we
suggest that barred spirals are more prone to circumnuclear star
formation because a higher fraction have $\Sigma_{nuc} > typical\
\Sigma_{crit}$.  We investigate the star formation activity in the
SONG sample further in the next section.

\section{Star Formation Rates, Gas Surface Densities \& the Schmidt Law}

In the left panel of Figure 2, we plot the star formation rate surface
density ($\Sigma_{SFR}$) against the molecular gas surface densities.
Above the typical threshold value of 100 \Msun\ pc$^{-2}$ (shown by
the vertical solid line), we find higher star formation rates but
there is considerable scatter in the data.  These galaxies, mostly
early-type barred bars, are indicated with the solid oval; they tend
to lie along the Schmidt law (Kennicutt 1998).  Below the critical
density galaxies have significantly lower star formation activity
(dashed oval) but it is worthwhile noting that these galaxies have
higher $\Sigma_{SFR}$ than would be predicted from the Schmidt law;
this behavior needs further analysis.  These data confirm the
hypothesis in the previous section that galaxies above the thresold,
i.e. barred spirals, are, on average, prone to higher star formation
activity.

\section{Central Concentrations}

In this section, we examine the re-distribution of the gas by the bar
by calculating a concentration parameter,
f$_{con}$=$\Sigma_{nuc}$/$\Sigma_{disk}$, introduced by Sakamoto et
al. (1999).  The results are plotted in the right panel of Figure 2.
The barred spirals have a mean f$_{con}$ of 60.22$\pm$13.6 whereas the
unbarred spirals have a mean f$_{con}$ of 21.27$\pm$7.32.  A K-S test
shows that the probability of the two samples being drawn from the
same distribution is 7\%.  The SONG data confirm the basic result of
Sakamoto et al. that barred spirals are more centrally concentrated
than unbarred spirals.  The two studies together provide solid
statistical evidence for bar-induced gas transport to the central kpc.

We also find that $\sim$10$^8$ \Msun\ of gas must have been
transported in by the bar, consistent with the OVRO-NRO data.
However, the mean difference in the mass between barred and unbarred
spirals in our sample is 1.7 $\times$10$^8$ \Msun, smaller than the
value derived by Sakamoto et al. because our sample was less biased
towards centrally concentrated galaxies.  As discussed in Sakamoto et
al. (1999), if the mass inflow rate in bars is on the order of 1
\Msun\ \yr, the two studies indicate that the bar destruction time
scale is at least 10$^8$ years, consistent with modeling studies.

\section{Bar Properties and the Central Concentration}
Parameters such as the bar length (or equivalently its pattern speed),
bar ellipticity, and the strength of the bulge are responsible for
controlling the rate of the mass inflow.  One expects that the mass
inflow rate is increased whenever the bulge is small relative to the
bar, the bar is thin, and the bar is long.
\subsection{Bar Length}

The bar length is the most straightforward parameter to calculate.  In
the left panel of Figure 3, we compare the deprojected bar length,
normalized by the galaxy diameter, to the central concentrations.  We
find that there is {\em no} correlation between the bar length and
f$_{con}$.  

Since bar length is a function of the Hubble type (e.g., Elmegreen \&
Elmegreen 1985), we checked the relationship between f$_{con}$
and bar length within each Hubble type bin and found no correlation.
Thus we conclude that bar length alone is a poor indicator of the
central mass concentrations in barred spirals.  Other factors such as
the bar ellipticity or the nuclear star formation rate are equally, if
not more, important in determining the central gas concentration.

\subsection{Bulge type}
 
\begin{figure}
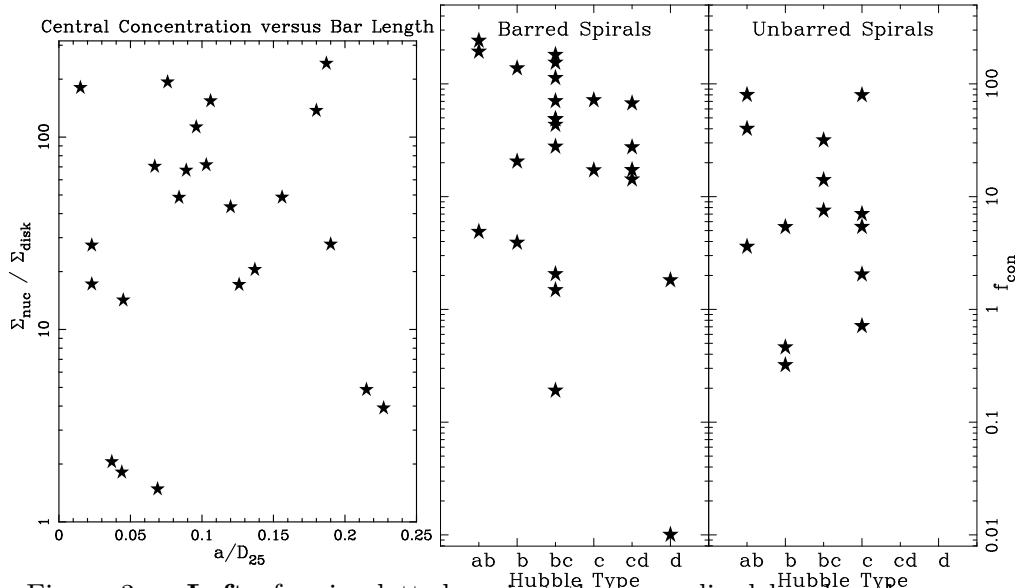

\vspace{0.8in}
\plotfiddle{cenconcvslength.ps}{2.7in}{0}{40}{40}{-210}{0}
\plotfiddle{cenconcvshub.ps}{2.7in}{0}{45}{45}{-46}{190}
\vspace{-3.7in}
\caption{{\bf Left:} f$_{con}$ is plotted against the normalized bar
lengths (a/D$_{25}$.  There is no correlation between the bar length
and the central concentration of gas, implying that bar ellipticity
and/or star formation are equally, if not more important.  {\bf
Right:} f$_{con}$ is plotted against the Hubble type for the barred
(left box) and unbarred spirals (right box).  There is a slight trend
of higher central concentrations in big bulges (i.e., earlier type
bars), but the trend is not statistically significant.}
\end{figure}

In the right hand panel of Figure 3 we compare f$_{con}$ to the bulge
size (using Hubble type as a proxy for bulge size).  For the barred
spirals (left box) we find that for each Hubble type
(ab:b:bc:c:cd:d) the mean value of the central concentration
(146:53:64:44:31:1) shows a trend of decreasing f$_{con}$ with later
Hubble types.  However there is large dispersion in the sample and a
least squares fit shows that the trend is not statistically
significant; a larger sample is necessary to confirm the trend.

If the central concentrations are higher in earlier type bars, then
the presence of inner Lindblad resonances may provide a satisfactory
explanation for the trend.  In models, the inflowing gas accumulates
in a circumnuclear ring located between the outer and inner ILR (Sheth
et al. 2000; Combes 1996).  Though the typical radius of such
circumnuclear rings is $\sim$ 1 kpc (e.g., Buta \& Combes 1996);
however, bigger bulges have deeper potentials and hence the ILR gas
ring is likely to be at smaller radii.

\section{Summary}

Using a 44 galaxy sample from SONG, we found that the central kpc of
bars have high molecular gas surface densities, and that a higher
fraction have $\Sigma_{nuc} > typical\ \Sigma_{crit}$; this may
explain the observed enhancement in circumnuclear star formation
activity in bars, particularly in early type bars.  We also find that
barred spirals are more centrally concentrated than unbarred spirals,
confirming the role of the bar in transporting gas inwards, consistent
with the OVRO-NRO study by Sakamoto et al. (1999).  The central
concentrations are weakly correlated with the bulge size, but not with
the bar length.  suggesting that ILRs, bar ellipticity and star
formation play equally important roles in determining the central kpc
gas properties.

\end{document}